\documentclass{ws-procs9x6}
\usepackage{graphicx}
\usepackage{color,epsfig}

\usepackage{amsmath}
\usepackage{amsfonts}
\usepackage{bm}

\def\drawbox#1#2{\hrule height#2pt
        \hbox{\vrule width#2pt height#1pt \kern#1pt
              \vrule width#2pt}
              \hrule height#2pt}

\def\Asym#1#2{\vcenter{\vbox{\drawbox{#1}{#2}
              \kern-#2pt       
              \drawbox{#1}{#2}}}}

\def\Acknowledgements{\bigskip  \bigskip {\begin{center}
              \bf Acknowledgments \end{center}}}

\newcommand {\beq} {\begin{equation}}
\newcommand {\eeq} {\end{equation}}
 \newcommand{\be}{\begin{eqnarray}}
\newcommand{\ee}{\end{eqnarray}}

\begin{document}

\title{Predictions for QCD from Supersymmetry
}

\author{F. Sannino
}

\address{NORDITA, Blegdamsvej 17, Copenhagen {\O} DK-2100, Denmark. }

%


\maketitle

\abstracts{We review the construction of the effective Lagrangians
of the Veneziano-Yankielowicz (VY) type for two non-supersymmetric
theories containing one Dirac fer\-mion in the two-index
antisymmetric or symmetric representation of the gauge group
(orientifold theories). Since these theories are planar equivalent,
at $N\to\infty$ to super Yang-Mill their effective Lagrangians
coincides with the bosonic part of the VY Lagrangian. We depart from
the supersymmetric limit in two ways. First, we consider finite but
still large values of $N$. Then $1/N$ effects break supersymmetry.
We suggest a minimal modification of the VY Lagrangian which
incorporates these $1/N$ effects, leading to a non-vanishing vacuum
energy density. We then analyze the spectrum at finite $N$. For
$N=3$ the two-index antisymmetric representation (one flavor) is
equivalent to one-flavor QCD. We show that in this case the scalar
quark-antiquark state is heavier than the corresponding pseudoscalar
state,  `` $\eta^{\prime}$''. Second, we add a small fermion mass
term. The fermion mass term breaks supersymmetry explicitly. The
vacuum degeneracy is lifted. The parity doublets split. We evaluate
the splitting. Finally, we include the $\theta$-angle and study its
implications.}

\section{Introduction}
\label{uno}

Recently it has been argued \cite{Armoni:2003gp,Armoni:2003fb} that
{\em non}-supersymmetric Yang-Mills theories with a fermion in the
two index symmetric or antisymmetric representation are
nonperturbatively equivalent to supersymmetric Yang-Mills (SYM)
theory at large $N$, so that exact results established in SYM theory
(e.g. \cite{{Shifman:1999mv},{Shifman:ia}}) should hold also in
these ``orientifold" theories. For example, the orientifold
theories, at large $N$, must have  an exactly calculable bifermion
condensate and an infinite number of degeneracies  in the spectrum
of color-singlet hadrons. The phenomenon goes under the name of
planar equivalence; it does not mean, however, the full
parent-daughter coincidence. For instance, at $N\rightarrow \infty$
the color-singlet spectrum of the orientifold theories does not
include composite fermions. The planar equivalence relates
corresponding {\em bosonic} sectors in the corresponding vacua of
the two theories. Some predictions for one-flavor QCD (which is the
antisymmetric orientifold daughter at $N=3$) were made along these
lines in \cite{Armoni:2003fb,ASV3}. Here we review the construction
of the effective Lagrangians for the orientifold field theories
which are able to capture relevant $1/N$ corrections
\cite{Sannino:2003xe}. Our starting point is the effective
Lagrangian for supersymmetric Yang-Mills.

 The name of {\it orientifold field theory} is borrowed
from string-theory terminology \cite{DiVecchia:2004ev}.

\section{Reviewing SYM  effective Lagrangian}
\label{due}


The effective Lagrangian for supersymmetric gluodynamics was found
by Veneziano and Yankielowicz (VY) \cite{Veneziano:1982ah}. In terms
of the composite color-singlet chiral superfield $S$,
\beq S= \frac{3}{32\pi^2 N }\,\mbox{Tr}\,W^2 \,, \eeq
it can be written as follows:
 \be { L}_{VY}&=& \frac{9\, N^2}{4\,\alpha}\, \int d^2\!\theta\,
d^2\!{\bar{\theta}}
\left(S S^{\dagger}\right)^{\frac{1}{3}} 
+\frac{N}{3} \int \! {\rm d}^2\theta\, \left\{ S  \ln \left(\frac{S
}{\Lambda^3}\right)^{N}-NS\right\}  + \mbox{H.c.} \, , \label{VY}
\nonumber \\\ee
where $\Lambda$ is a parameter related to the fundamental SYM scale
parameter \footnote{The Grassmann integration is defined in such a
way that $\int \, \theta^2\, d^2\theta =2$.}. We singled out the
factor $N^2$ in the K\"{a}hler term to make the parameter $\alpha$
scale as $N^0$, see Eq.~(\ref{alphascaling}) below.
%
%
With our definitions, the gluino condensate scales as $N$.

Requiring the mass of the excitations to be $N$ independent one
deduces \beq \alpha \sim {N^0} \,. \label{alphascaling} \eeq
Indeed, the common mass of the bosonic and fermionic components of
$S$ is $M=2\alpha\, \Lambda /3$.  The chiral superfield $S$ at the
component level has the standard decomposition $S(y)=\varphi(y) +
\sqrt{2} \theta \chi(y) + \theta^2 F(y)$, where $y^\mu$ is the
chiral coordinate, $y^\mu=x^\mu - i \theta \sigma^\mu \bar{\theta}$,
and
\beq \varphi\ ,\quad \sqrt{2}\chi\ ,\quad F =\frac{3}{64\pi^2
N}\times \, \left\{
\begin{array}{l}
-\lambda^{a,\alpha}\lambda^{a}_{\alpha}\\[3mm]
G^a_{\alpha\beta}\lambda^{a,\beta} +2i D^a \lambda^{a}_{\alpha} \\[3mm]
-\frac{1}{2} G^a_{\mu\nu} G^{a\mu\nu}+ \frac{i}{2}G^a_{\mu\nu}
\tilde{G}^{a\mu\nu}+\mbox{f.t.}
\end{array}
\right. \label{decomp} \eeq
where f.t. stands for fermion terms.

The complex field $\varphi$ represents the scalar and pseudoscalar
gluino-balls while $\chi$ is their fermionic partner. It is
important that the $F$ field {\em must be treated as auxiliary}.


This lagrangian is not complete as pointed out in
\cite{Kovner-Shifman}. Recently an extended VY Lagrangian which
passes a number of non trivial consistency checks while naturally
yielding the VY effective theory augmented by the missing terms
pointed out in \cite{Kovner-Shifman} has been constructed
\cite{Merlatti:2004df}. Such an extension requires the introduction
of a glueball superfield, i.e. a chiral superfield with zero $R$
charge. Earlier attempts of generalizing the VY Lagrangian
containing glueball degrees of freedom are discussed in the
literature \cite{{Shore:1982kh},{Kaymakcalan:1983jh},Farrar:1998rm,
Sannino:1997dd}. These extensions were triggered, in part,  by
lattice simulations of SYM spectrum \cite{Feo:2002yi}. Since in this
paper we are interested in the mesonic degrees of freedom we will
not consider the generalized version of the VY
\cite{Merlatti:2004df} although it is now straightforward to
generalized our results to the improved VY theory.

For our purposes of most importance are the scale and chiral
anomalies,
\begin{equation}
\partial^\mu J_\mu = \frac{N }{16\pi^2}\,
G_{\mu\nu}^a\tilde{G}^{a,\, \mu\nu}\, ,\quad J_\mu= -
\frac{1}{g^2}\, \lambda^a \sigma_\mu \,\bar\lambda^a\ , \quad
%
%
\vartheta^\mu_{\mu}=- \frac{3 N }{32 \pi^2}\, G_{\mu\nu}^a
{G}^{a ,\, \mu\nu}\,\, , \eeq
where $J_\mu$ is the chiral current and $\vartheta^{\mu\nu}$ is the
standard (conserved and symmetric) energy-momentum tensor.

In SYM theory these two anomalies belong to the same supermultiplet
\cite{Ferrara-Zumino} and, hence, the coefficients are the same (up
to a trivial 3/2 factor due to normalizations). In the orientifold
theory the coefficients of the chiral and scale anomalies coincide
only at $N=\infty$; the subleading terms are different.

Summarizing, the component form of the VY Lagrangian is
\beq {L}_{\rm VY}=\frac{N^2}{\alpha}\left(\varphi\, \bar\varphi
\right)^{-2/3}\,
\partial_\mu\bar\varphi\,\partial^\mu\varphi-\frac{4\, \alpha\,
N^2}{9}\, \left(\varphi\, \bar\varphi \right)^{2/3}
\ln\bar\varphi\,\ln\varphi +\mbox{fermions}\,, \label{vycomponent}
\eeq
where we set $\Lambda =1$ to ease the notation.

\section{Effective Lagrangians in orientifold theories}
\label{tre} In the theory with the fermions in the two
index-antisymmetric representation the trace and the chiral
anomalies are
\begin{eqnarray}
\vartheta^{\mu}_{\mu} &=&2N\left[N + \frac{4}{9}\right]\left(F +
\bar F\right) = -3\left[N +
\frac{4}{9}\right]\frac{1}{32\pi^2}\, G_{\mu\nu}^a {G}^{a ,\, \mu\nu} \ , \label{trace}\\[3mm]
\partial^{\mu} J_{\mu}
&=&i\,\frac{4N}{3}\,\left[ N - 2\right]\, \left(\bar F - F\right)=
\left[N - 2\right]\frac{1}{16\pi^2}\, G_{\mu\nu}^a {\tilde{G}}^{a
,\, \mu\nu} \ , \label{axial}
\end{eqnarray}
where
\begin{eqnarray}
\varphi= -\frac{3}{32\pi^2\, N}\,
 \widetilde{\psi}^{\alpha,[i,j]} \psi_{\alpha,[i,j]} \ ,
\label{phi1}
\end{eqnarray}
and $F$ is given in Eq.~(\ref{decomp}). The gluino field of
supersymmetric gluodynamics is replaced in this theory by two Weyl
fields, $ \widetilde{\psi}^{\alpha,[i,j]}$ and
$\psi_{\alpha,[i,j]}$, which can be combined into one Dirac spinor.
The color-singlet field $\varphi$ is now bilinear in $
\widetilde{\psi}^{\alpha,[i,j]}$ and  $\psi_{\alpha,[i,j]}$. Note
the absence of the color-singlet fermion field $\chi$ which was
present in supersymmetric gluodynamics.

In the infinite $N$ limit we will deal with the same coefficient in
both anomalies, much in the same way as in SUSY gluodynamics. In
fact, in this limit the boson sector of the daughter theory is
identical to that of the parent one \cite{Armoni:2003gp}, and,
hence,  the effective Lagrangian must have  exactly the same form as
in Eq. (\ref{vycomponent}), with the omission of the fermion part
and the obvious replacement of $\lambda^a \lambda^a$ by $2\,
 \widetilde{\psi}^{\alpha,[i,j]} \psi_{\alpha,[i,j]}\,$
in the definition of $\varphi$. The dynamical degrees described by
this Lagrangian are those related to $\varphi$, i.e. scalar and
pseudoscalar quark mesons. Hence we recover all
supersym\-metry-based bosonic properties such as degeneracy of the
opposite-parity mesons. Moreover, in this approximation the vacuum
energy vanishes. In the following we will concentrate on the two
index antysimmetric representation. The analysis for the two index
symmetric is presented in \cite{Sannino:2003xe}. Recently theories
with fermions in higher representations --in particular the two
index symmetric representation-- were shown to play a relevant role
when used as the underlying strong dynamics triggering electroweak
symmetry breaking \cite{Sannino:2004qp,Hong:2004td}.

\subsection{Effective Lagrangians in the orientifold
theories at finite $N$} \label{tre-finite-N}

The effective Lagrangians approach turns to be very useful since it
is hard to compute $1/N$ corrections in the underlying theory.

What changes must be introduced at finite $N$? First of all, the
overall normalization factor $N^2$ in Eq.~(\ref{vycomponent}) is
replaced by some function $f(N)$ such that $f(N)\to N^2$ at
$N\to\infty$. Moreover, the anomalous dimension of the operator $
\widetilde{\psi}^{\alpha,[i,j]} \psi_{\alpha,[i,j]}\,$ no more
vanishes. In fact, the renormalization-group invariant combination
is \beq \left(N\, g^2\right)^\delta\,
\widetilde{\psi}^{\alpha,[i,j]} \psi_{\alpha,[i,j]} \,, \qquad
\delta \equiv \frac{\left(1-\frac{2}{N}\right)
\left(1+\frac{1}{N}\right)}{\left(1+\frac{4}{9N}\right)}
-1\approx-\frac{13}{9N}\,. \label{exanomdi} \eeq It is just this
combination that should enter in the definition of the variable
$\varphi$ replacing Eq.~(\ref{phi1}), \beq \varphi=
-\frac{3}{32\pi^2\, (N-2)}\, \left(N\, g^2\right)^\delta\,
  \widetilde{\psi}^{\alpha,[i,j]} \psi_{\alpha,[i,j]} \,. \label{exformu} \eeq
In passing from Eq.~(\ref{phi1}) to Eq.~(\ref{exformu}), in addition
to taking account of the anomalous dimension, we replaced $N$ in the
denominator by $N-2$. The distinction between these two factors is a
subleading $1/N$ effect. The physical motivation for the above
replacement is as follows.  At $N=2$ the antisymmetric quark field
looses color, and, thus, $\langle \widetilde{\psi}^{\alpha,[i,j]}
\psi_{\alpha,[i,j]}  \rangle$ must vanish. The definition (\ref
{exformu}) guarantees, that it does vanish.


As constraints we require: (a) the finite-$N$ effective Lagrangian
to recover (\ref{vycomponent}) once the $1/N$ corrections are
dropped and (b) the scale and chiral anomalies (\ref{trace}),
(\ref{axial}) to be satisfied. The first requirement means, in
particular, that we continue to build ${ L}_{\rm eff}$ on a single
(complex) dynamical variable $\varphi$. Equations (\ref{trace}) and
(\ref{axial}) tell us that we cannot maintain the ``supersymmetric''
structure of the potentail term. We {\em have} to ``untie'' the
chiral and conformal dimensions of the fields in the logarithms, see
Eq. (\ref{vycomponent}). They cannot be just powers of $\varphi$
since in this case the chiral and conformal dimensions would be in
one-to-one correspondence, and the coefficients of the chiral and
scale anomalies would be exactly the same, modulo the normalization
factor 3/2. At this stage a non-holomorphicity must enter the game.

Let us introduce the fields
\beq \Phi =
\varphi^{1+\epsilon_1}\,\bar\varphi^{-\epsilon_2}\,,\qquad \bar\Phi
= \bar\varphi^{1+\epsilon_1}\,\varphi^{-\epsilon_2}\,,
\label{newfields} \eeq where $\epsilon_{1,2}$ are parameters
$O(1/N)$, \beq \epsilon_1 =- \frac{7}{9\,N}\,,\qquad \epsilon_2
=-\frac{11}{9\,N}\,. \label{newparam} \eeq
The scale and chiral dimensions of $\bar\Phi$ and $\Phi$ are such
that using $\bar\Phi$ and $\Phi$ in the logarithms, we will solve
the problem of distinct $1/N$ corrections in the coefficients of the
scale  and chiral anomalies. The above replacement (\ref{newfields})
is minimal (see \cite{Sannino:2003xe}). An important point to recall
here is that

 ($\star$) our replacement does not spoil the fact that $G^2$ and
$G\tilde{G}$ are real and imaginary parts of a certain field. (The
operators $G^2$ and $G\tilde{G}$ will be identified through the
non-invariance of the Lagrangian under the scale and chiral
transformations, see below.)

The vacuum expectation value (VEV) of $G\tilde{G}$ vanishes in any
gauge theory with massless fermion field while this is not the case
for the vacuum expectation value of $G^2$, which develops a VEV at
the subleading order in $1/N$. Preserving the property ($\star$)
above leaves open a single route: the $O(1/N)$ term to be added to
${ L}_{\rm eff}$ which will give rise to $\langle G^2\rangle$ must
be scale invariant by itself.

As for the kinetic term to begin with, we will make the simplest
assumption and leave the kinetic term the same as in ${L}_{\rm VY}$.
Other choices do not alter the overall picture in the qualitative
aspect.

The effective Lagrangian in the finite-$N$ orientifold theory reads:
\beq
 { L}_{\rm eff}=f(N)\left\{
\frac{1}{\alpha}\left(\varphi\, \bar\varphi \right)^{-2/3}\,
\partial_\mu\bar\varphi\,\partial^\mu\varphi-\frac{4\alpha}{9}\,
\left(\varphi\, \bar\varphi \right)^{2/3}\,\left(
\ln\bar\Phi\,\ln\Phi - \beta \right)\right\}\,, \label{fnocomponent}
\eeq where $\beta $ is a numerical (real) parameter, $\beta =
O(1/N)\,$,  and \beq f(N) \to N^2 \,\,\, \mbox{at}\,\,\,
N\to\infty\,. \eeq
The variations of this effective action under the scale and chiral
transformations (i.e. $\varphi \to (1+3\gamma)\varphi $ and $\varphi
\to (1+2i \gamma)\varphi $, respectively, with real $\gamma$) are
\begin{eqnarray}
\delta { S}_{\rm eff}^{\rm scale}&=& \int d^4x\left\{-4\,
\frac{\alpha\, f}{3}\left(\varphi\, \bar\varphi \right)^{2/3}\left(
1+\epsilon_1 -\epsilon_2\right)
\left(\ln\bar\Phi +\ln \Phi\right)\right\}\,,\nonumber\\[4mm]
\delta { S}_{\rm eff}^{\rm chiral}&=& \int d^4 x\left\{-8i\,\,
\frac{\alpha\, f}{9}\left(\varphi\, \bar\varphi \right)^{2/3}\left(
1+\epsilon_1 +\epsilon_2\right) \left(\ln\bar\Phi -\ln
\Phi\right)\right\}\,, \label{variat}
\end{eqnarray}
where the parameters $\epsilon_{1,2}$ are defined in
Eq.~(\ref{newparam}). Comparing with Eqs.~(\ref{trace}) and
(\ref{axial}) we conclude that
\begin{eqnarray}
G_{\mu\nu}^a {G}^{a ,\, \mu\nu} &\propto& -N\left(\varphi\,
\bar\varphi
\right)^{2/3}\left(\ln\bar\Phi +\ln \Phi\right)\,,\nonumber\\[3mm]
G_{\mu\nu}^a \tilde{G}^{a ,\, \mu\nu} &\propto& -N\,i\,
\left(\varphi\, \bar\varphi \right)^{2/3}\left(\ln\bar\Phi -\ln
\Phi\right)\,. \label{indenti}
\end{eqnarray}
Minimizing the potential term in the Lagrangian (\ref{fnocomponent})
we find that the minimum occurs at \beq \ln \varphi =
\frac{2}{3}\,\beta +O(1/N^2)\,, \label{polmi} \eeq and the minimal
value of the potential energy --- i.e. the vacuum energy density ---
is
\beq V_{\rm min}={ E}_{\rm vac} = -\frac{4\alpha\, f}{9}\,\beta +
O(N^0)\,. \label{vmin}
 \eeq
Here the value of ${ E}_{\rm vac}$ is determined by the
non-logarithmic term in the potential energy. The logarithmic term
enters only at the level $O(N^0)$.

There is a very important self-consistency check. One can
alternatively define the vacuum energy density as
$\frac{1}{4}\langle \vartheta^\mu_\mu\rangle $, where the trace of
the energy momentum tensor is in turn proportional to $G_{\mu\nu}^a
{G}^{a ,\, \mu\nu}$, see Eq.~(\ref{indenti}). In this method ${
E}_{\rm vac}$ will be determined exclusively by the logarithmic
term. In fact, it is not difficult to see that
\begin{eqnarray}
{ E}_{\rm vac} &=& -\frac{3N + {4}/{3}}{128\pi^2}\left\langle
G_{\mu\nu}^a
G^{a,\,\mu\nu}\right\rangle\nonumber\\[3mm]
&=& - \frac{\alpha\, f}{3}\, \left\langle \ln\bar\Phi +\ln
\Phi\right\rangle  + O(N^0) =-\frac{4\alpha\, f}{9}\,\beta +
O(N^0)\,, \label{altt}
\end{eqnarray}
in complete agreement with Eq.~(\ref{vmin}).

{}From the above consideration it is clear that the (infrared part
of the) vacuum energy density is negative {\em if} $\beta >0$. In
\cite{Sannino:2003xe} we argued that this is indeed the case.

\subsection{Lifting the spectrum degeneracy at finite $N$ and gluino mass $m$.}
\label{tre-finiteN-spectrum}

At $N\to\infty$ the orientifold theory inherits from its
supersymmetric parent an infinite number of degeneracies in the
bosonic spectrum. At the effective Lagrangian level this property
manifests itself in the degeneracy of the scalar/pseudoscalar
mesons. At finite $N$ we expect this degeneracy to be lifted by
$1/N$ effects as well as the explicit presence of a gluino mass.

The leading $1/N$ and $m$ corrections can be considered
simultaneously using the Lagrangian found in \cite{Sannino:2003xe}:
\begin{eqnarray} &&f(N)\left\{ \frac{1}{\alpha}\left(\varphi\, \bar\varphi
\right)^{-2/3}\,
\partial_\mu\bar\varphi\,\partial^\mu\varphi-\frac{4\alpha}{9}\,
\left(\varphi\, \bar\varphi \right)^{2/3}\,\left(
\ln\bar\Phi\,\ln\Phi - \beta \right)\right\}
\nonumber \\
&& + {{\frac{4\, m\,N(N-2)}{3\lambda
\left(8\pi^2\,\lambda\right)^{\delta}}}} \left( \varphi +
\bar{\varphi}\right)\ . \nonumber \\\label{fnocomponent+mass}
\end{eqnarray}
%

%

To explore the scalar/pseudoscalar splitting one must study
excitations near the vacuum in the Lagrangian (\ref{fnocomponent}).
Let us define
 \begin{eqnarray}
\varphi = \langle \varphi \rangle_{\rm vac}\left(1
 + a\,h\right)\ ,  \qquad
 h=\frac{1}{\sqrt{2}} \left(\sigma + i\,\eta^{\prime}
\right) \, ,
\end{eqnarray}
$\sigma$ and $\eta^{\prime}$ are two real fields and $a$ is a
constant which is determined by requiring the standard normalization
of the kinetic term for the complex field $h$,
\begin{eqnarray}
a^2=\frac{\alpha}{f} \, |\langle  \varphi \rangle|^{{-\frac{2}{3}}}
\ ,
\end{eqnarray}
The vacuum expectation value reads:
\begin{eqnarray}
\langle \varphi \rangle = \Lambda^3 \left(1 + \frac{2}{3}\beta +
\frac{3m}{\alpha \lambda \Lambda}\right ) + {
O}\left(m^2,N^{-2},mN^{-1}\right)\ ,
\end{eqnarray}
yielding the following vacuum energy density:
\begin{eqnarray}
{ E}_{\rm vac}=V_{\rm min} = -\frac{4\alpha f}{9} \beta \Lambda^4 -
\frac{8N^2}{3\lambda} m\Lambda^3 +{ O}\left(m^2,N^{0},mN\right) \ .
\end{eqnarray}
For the spectrum we predict the following ratio of the pseudoscalar
to scalar mass:
%
%
 \begin{eqnarray}
\frac{M_{\eta^{\prime}}}{M_{\sigma}} = 
 1 -\frac{22}{9N} -\frac{4}{9}\beta -  \frac{m}{\alpha \lambda \Lambda}+
{O}(m^2,N^{-2},mN^{-1}) \ .\label{spectrum-ration-Nm}
\end{eqnarray}
The gluon condensate is:
\begin{eqnarray}
 \frac{\langle G^a_{\mu\nu}G^{a,\mu\nu} \rangle}{64\pi^2} =
\frac{4\, N \, m}{3\lambda}\, \Lambda^3 +
\frac{8}{27}\alpha\,N\beta\Lambda^4+ { O}\left(m^2,N^{-1},mN^0
\right)\ .
\end{eqnarray}
These results show that the contribution of the fermion mass
reinforces the effect of the finite $N$ contribution. Interestingly
the scalar state becomes even more massive than the pseudoscalar
state when considering finite both $N$ and $m$.

The $\theta$-angle dependence of the vacuum energy for the fermions
in the two-index antisymmetric representation of the gauge group is
\begin{eqnarray} { E}_{\rm vac} =
\frac{8N^2}{3\,\lambda}\, m \Lambda^3\, {\rm
min}_{k}\left\{-\cos\left[\frac{\theta + 2\pi\,k
}{N-2}\right]\right\}  -\frac{4\alpha f}{9} \beta \Lambda^4 \
.\end{eqnarray}
The $N-2$-fold vacuum degeneracy is lifted due to the presence of a
mass term in the theory, yielding a unique vacuum.

%

As was mentioned, for $N=3$  the two-index antisymmetric
representation is equivalent to one-flavor QCD (with a Dirac fermion
in the fundamental representation). We then predict that in
one-flavor QCD the scalar meson made of one quark and one anti-quark
is heavier than the pseudoscalar one (in QCD the latter is
identified with the $\eta^{\prime}$ meson).

\section{Conclusions}
\label{sec:Discussion}

We constructed  the effective Lagrangians of the
Veneziano-Yankielowicz type for orientifold field theories, starting
from the underlying  SU$(N)$ gauge theory with the Dirac fermion in
the two-index antisymmetric (symmetric) representation of the gauge
group. These Lagrangians incorporate ``important" low-energy degrees
of freedom (color singlets) and implement (anomalous) Ward
identities. At $N\to\infty$ they coincide with the bosonic part of
the VY Lagrangian.
 The orientifold effective
Lagrangians at $N=\infty$ display the vanishing of the cosmological
constant and the spectral  degeneracy (i.e. the scalar-pseudoscalar
degeneracy).

The most interesting question we addressed is the
finite-$N$/finite-$m$ generalization. To the leading order in $1/N$
we demonstrated the occurrence of a negative vacuum energy density,
and of the gluon  condensate. We first derived these results at
$m=0$ and then extended them to include the case $m\neq 0$. At $N=3$
the theory with one  Dirac fermion in the two-index antisymmetric
representation of the gauge group is in fact one-flavor QCD. Our
analysis of the finite-$N$ effective Lagrangian illustrates the
emergence of the gluon condensate in this theory. The vacuum
degeneracy typical of supersymmetric gluodynamics does not disappear
at finite $N$. However, introduction of mass $m\neq 0$ lifts the
vacuum degeneracy, in full compliance with the previous
expectations. Both effects, $N\neq \infty$ and $m\neq 0$ conspire to
get lifted the scalar-pseudoscalar degeneracy. We evaluated the
ratio $M_{\eta^{\prime}}/M_{\sigma}$. Finally, we studied the
effects of finite $\theta$, as they are exhibited in the orientifold
effective Lagrangian. In the diagram presented in figure we sketch
how the different theories can be related as function of the mass of
the gluino and $1/N$ corrections. \begin{figure}[t]
  \begin{center}
\mbox{\kern 1.cm \epsfig{file=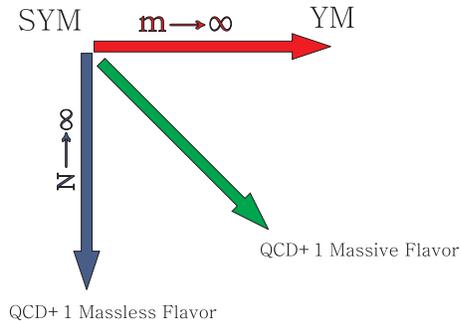,width=6.0true
cm,angle=0}}
  \end{center}
  \caption{Schematic diagram which summarizes the link between different gauge theories.}
\end{figure}%
\vskip -.5cm \Acknowledgements It is a pleasure for me to thank M.
Shifman for careful reading of the manuscript and for a very
enjoyable collaboration.

\end{document}